\documentclass[prl,twocolumn,notitlepage,showpacs,superscriptaddress]{revtex4-1}
\usepackage[left=1.75cm,right=1.75cm,top=1.85cm,bottom=1.9cm]{geometry}
\usepackage{amsmath}
\usepackage{amssymb}
\usepackage{graphicx}
\usepackage{epstopdf}
\usepackage{subfigure}

\renewcommand{\k}{\mathbf{k}}
\renewcommand{\r}{\mathbf{r}}
\newcommand{\q}{\mathbf{q}}
\newcommand{\nablaK}{\boldsymbol{\nabla}_k}

\newcommand{\BZ}{\textrm{BZ}}

\newcommand{\spacer}{\vphantom{\frac{1}{2}}}

\newcommand{\BerryConn}{\mathbf{A}}

\newcommand{\HamConf}{\overline{V}}

\newcommand{\Kummer}[3]{ {}_1F_{1}\left[\begin{smallmatrix} #1 \\#2 \end{smallmatrix};#3\right] }

\newcommand{\PI}{\hat{\pi}^{\vphantom{\dag}}}
\newcommand{\PID}{\hat{\pi}^\dag}
\newcommand{\PII}{\hat{\boldsymbol{\Pi}}}

\usepackage{color}

\DeclareMathOperator{\tr}{tr}

\newcommand{\TOp}{\hat{\mathcal{T}}}

\newcommand{\Ham}{\hat{H}}

\newcommand{\Ch}{\mathcal{C}}
\newcommand{\ketbra}[2]{\left|#1\middle>\middle<#2\right|}
\newcommand{\braket}[2]{\left<#1\middle|#2\right>}
\newcommand{\braOPket}[3]{\left<#1\middle|#2\middle|#3\right>}
\newcommand{\bra}[1]{\left<#1\right|}
\newcommand{\ket}[1]{\left|#1\right>}
\newcommand{\OPc}[2]{\hat{#1}_{#2}^{\dag}}
\newcommand{\OP}[2]{\hat{#1}_{#2}^{\vphantom{\dag}}}
\newcommand{\CD}[1]{\OPc{c}{#1}}
\newcommand{\C}[1]{\OP{c}{#1}}
\newcommand{\A}[1]{\OP{a}{#1}}
\newcommand{\AD}[1]{\OPc{a}{#1}}

\newcommand{\E}{\epsilon}

\newcommand{\dens}{\hat{\rho}}

\newcommand{\comm}[2]{\left[#1,~#2\right]}

\begin{document}
\title{Position-Momentum Duality and Fractional Quantum Hall Effect in Chern Insulators}
\author{Martin Claassen}
 \affiliation{Department of Applied Physics, Stanford University, CA 94305, USA} 
 \affiliation{Stanford Institute for Materials and Energy Sciences, SLAC \& Stanford University, CA 94025, USA}
\author{Ching Hua Lee}
 \affiliation{Department of Physics, Stanford University, CA 94305, USA} 
\author{Ronny Thomale}
 \affiliation{Institute for Theoretical Physics and Astrophysics, University of W\"urzburg, D 97074 W\"urzburg}
\author{Xiao-Liang Qi}
 \affiliation{Department of Physics, Stanford University, CA 94305, USA} 
\author{Thomas P. Devereaux}
 \email[Author to whom correspondence should be addressed to: M. C. (\href{mailto:mclaassen@stanford.edu}{mclaassen@stanford.edu})
]{}
\affiliation{Stanford Institute for Materials and Energy Sciences, SLAC \& Stanford University, CA 94025, USA}
\begin{abstract}
We develop a first quantization description of fractional Chern insulators that is the dual of the conventional fractional quantum Hall (FQH) problem, with the roles of position and momentum interchanged. In this picture, FQH states are described by anisotropic FQH liquids forming in momentum-space Landau levels in a fluctuating magnetic field. 
The fundamental quantum geometry of the problem emerges from the interplay of single-body and interaction metrics, both of which act as momentum-space duals of the geometrical picture of the anisotropic FQH effect. We then present a novel broad class of ideal Chern insulator lattice models that act as duals of the isotropic FQH effect. The interacting problem is well-captured by Haldane pseudopotentials and affords
a detailed microscopic understanding of the interplay of interactions and non-trivial quantum geometry.
\end{abstract}
\maketitle

The effects of topology and quantum geometry in condensed matter physics have recently garnered immense attention. With topological insulators and the quantum anomalous Hall effect constituting the well-understood case of non-interacting or weakly-correlated electron systems \cite{Haldane1987, HasanKaneRMP, QiZhangRMP}, the recent theoretical discovery of the fractional quantum Hall (FQH) effect in nearly-flat bands with non-trivial topology \cite{Mudry2011, ShengNature2011, RegnaultPRX2011, BergholtzReview2013, ParameswaranReview2013} poses deep questions regarding the confluence of strong interactions and non-trivial quantum geometry. On one hand, an experimental realization of such a fractional Chern insulator (FCI) could conceivably push relevant energy scales by an order of magnitude, paving the way to robust FQH signatures \cite{XiaoOkamotoNature2011, VenderbosPRL2011, KourtisPRB2012, CooperPRL2012, LiuOrganometallicPRL2013,GrushinPRL2014}. On the other hand, the disparity of conventional Landau levels and flat bands with non-zero Chern number $\Ch$ suggests a rich playground to realize novel states with topological order that cannot be attained in a conventional electron gas in a magnetic field \cite{BarkeshliQiPRX2012,WuRegnaultBernevigPRL2013}, while simultaneously presenting a profound challenge to understand the underlying microscopics of strong interactions.

Most of our current understanding regarding FCIs stems from exact diagonalization (ED) of small clusters for $\Ch=1$ \cite{WangPRL2011,WuPRB2011,BernevigTranslationalSymmetriesPRB2012, LiuRepellinPRB2013,LaeuchliPRL2013,GrushinPRB2012,KourtisPRL2014,GrushinARXIV2014, LiuKovrizhinPRB2012,LuoChenPRB2013, WuJainSunPRB2012, ScaffidiMoellerPRL2012,LiuBergholtzPRB2013, WuRegnaultBernevigPRB2014} and $\Ch>1$ \cite{LiuPRL2012, WangYaoPRL2012, SterdyniakPRB2013, LiuPRB2013, WangYaoPRB2012}, mutatis-mutandis mappings of the Hilbert space of flat Chern bands to the lowest Landau level (LLL) \cite{QiPRL2011, BarkeshliQiPRX2012, ChinghuaPRB2013, ChinghuaPRB2014, ChaomingPRB2013, WuRegnaultBernevigPRB2012} or vice versa \cite{WuRegnaultBernevigPRL2013}, and approximate long-wavelength projected density algebra \cite{ParameswaranPRB2012, EstiennePRB2012, RoyARXIV2012, RoyARXIV2014, GoerbigEPJ2012, MurthyShankarPRB2012,DobardzicPRB2013, RepellinPRB2014}. The latter approaches however treat exclusively the universal long-wavelength continuum limit of the FQH problem, whereas the presence and relevance of the lattice manifests itself in the short-wavelength physics. This conundrum is highlighted by the zoo of FCI lattice models established so far, which display strongly varying proclivities to host stable FQH phases that do not correlate well with simple measures such as ``flatness'' of band dispersion and Berry curvature.

At its heart, the theoretical challenge stems from the fact that the immense success in describing the microscopics of the conventional FQH effect resists a simple description in second quantization \cite{OrtizSeidel2013} that is essential to describe interacting electrons on the lattice \cite{SunPRL2011,TangWenPRL2011,HuPRB2011,WangRanPRB2011,YangHigherCPRB86, TrescherPRB2012}. A resolution is crucial to provide a foundation for studies of non-Abelian phases \cite{MooreRead1991,ReadRezayi1999,Ardonne1999} and to provide microscopic insight that can ultimately drive experimental discovery.

In this work, we develop a first quantization description of FQH states in FCIs that leads to an effective Hamiltonian that is the dual of the usual FQH problem, with the roles of position and momentum interchanged. In this picture, FCI analogues of FQH states are described by anisotropic FQH liquids forming in momentum-space Landau levels in a fluctuating magnetic field.
The challenge of understanding FQH states in FCIs reduces to a variational problem of determining the deformation of the guiding-center orbitals due to the presence of the lattice, in analogy to Haldane's geometrical picture of the anisotropic FQH effect (FQHE) \cite{HaldanePRL2011, YangPRB2012, QiuPRB2012}.
Guided by these insights, we then present and provide examples of a broad class of ideal FCI host lattice models that constitute FCI analogs of the \textit{isotropic} FQHE. These models afford a particularly simple description of the interacting low-energy dynamics, acting as FQH parent Hamiltonians with emergent guiding center and SU($\Ch$) symmetry. The effects of quantum geometry and Berry curvature fluctuations are analyzed in terms of Haldane pseudopotentials.

Consider a 2D band insulator hosting an isolated fractionally-filled flat band with non-zero $\Ch$, generically described by an $N$-orbital Bloch Hamiltonian $\hat{\mathbf{h}}_\k$. In band basis, the flat band of interest is spanned by Bloch states $\ket{u_\k}$ with dispersion $\hat{\mathbf{h}}_\k \ket{u_\k} = \E \ket{u_\k}$. If the band gap is larger than intra-band interactions, then the kinetic energy is effectively quenched while momentum-dependent orbital mixing for $\ket{u_\k}$ gives rise to a non-trivial quantum geometry, expressed by a gauge field and a Riemann metric on $\mathbb{C}\mathbb{P}^{N-1}$ for the Bloch band, the Berry curvature $\Omega(\k)$ and Fubini-Study metric $g_{\mu\nu}(\k)$:
\begin{align}
	\Omega(\k) &= \epsilon^{\mu\nu} \partial_{k_\mu} \BerryConn_\nu(\k) \hspace{0.8cm} \BerryConn_\nu(\k) = -i\braket{u_\k}{\partial_{k_\nu} u_\k} \\
g_{\mu\nu}(\k) &= \tfrac{1}{2}\braOPket{\partial_{k_\mu} u_\k}{\left[1 - \ketbra{u_k}{u_k}\right]}{\partial_{k_\nu} u_\k} + \left(\mu\leftrightarrow\nu\right)
\end{align}
Here, $\BerryConn_\nu(\k)$ is the Berry connection, and $\Ch = \frac{1}{2\pi}\int_{\rm BZ} d^2k~ \Omega(\k)$.
We use lattice constants $a_0 = \hbar = 1$.

The first task at hand is to describe the guiding-center basis. In the case of an isotropic free electron gas in a magnetic field, Laughlin constructed a series of incompressible FQH trial wave functions \cite{Laughlin1983}
for odd-fraction filling factors, from a single-body basis of radially-localized symmetric-gauge LLL wave functions $\braket{\r}{m} \sim z^m e^{-|\r|^2/4}$, which are uniquely determined by demanding that they are eigenstates of the angular momentum operator. While angular momentum does not readily translate to the lattice, a key observation is that such states are simultaneous eigenstates of a parabolic confinement potential $\hat{V}(\mathbf{r}) = \frac{\lambda}{2} \r^2$ projected to the LLL: $ \left[\sum_{m'} \ketbra{m'}{m'}\right] \hat{V}(\r) \ket{m} = \lambda \left(m+1\right) \ket{m}$.

A natural way to adapt this construction to an FCI is to consider \textit{anisotropic} confinement on a lattice, as a tool to determine the guiding-center basis:
\begin{equation}
	\hat{V}(\r)= \tfrac{1}{2} \lambda~ x_{\mu} \eta^{\mu\nu} x_{\nu} \label{eq:ConfinementPotential}
\end{equation}
Here, $\eta^{\mu\nu}$ is a unimodular Galilean metric which \textit{a priori} serves as a variational degree of freedom, constrained to retain the discrete rotational symmetries of the host lattice, and $\r = m_1 \mathbf{a}_1 + m_2 \mathbf{a}_2, m_{1,2} \in \mathbb{Z}$ indexes the unit cell with lattice vectors $\mathbf{a}_{1,2}$. 
Placing $\hat{V}$ on a $L \times L$ lattice via appropriate long-distance regularization \cite{SUPPLEMENTARYdetailedderivation}, the low-energy dynamics follow from projection onto the flat band with projector $\hat{P} = \sum_\k \ketbra{\k}{u_\k}$, and taking $L \to \infty$:
\begin{align}
	\HamConf =  \hat P\hat V(\r)\hat P^\top= \frac{\lambda}{2} \PII_\mu \eta^{\mu\nu} \PII_\nu + \frac{\lambda}{2} \eta^{\mu\nu} g_{\nu\mu} \hspace{0.27cm} \k \in {\rm BZ}        \label{eq:confinementHamiltonian0}
\end{align}
where $\PII_\mu$ are momentum-space analogues to the usual canonical momentum operators, with $\mu = x,y$:
\begin{align}
	\PII_\mu = -i\partial_{k_\mu} + A_\mu(\k) \hspace{0.8cm} [\PII_\mu,~\PII_\nu] = -i \epsilon_{\mu\nu}~ \Omega(\k)
\end{align}

\begin{figure}
\centering
	\includegraphics[width=8.9cm]{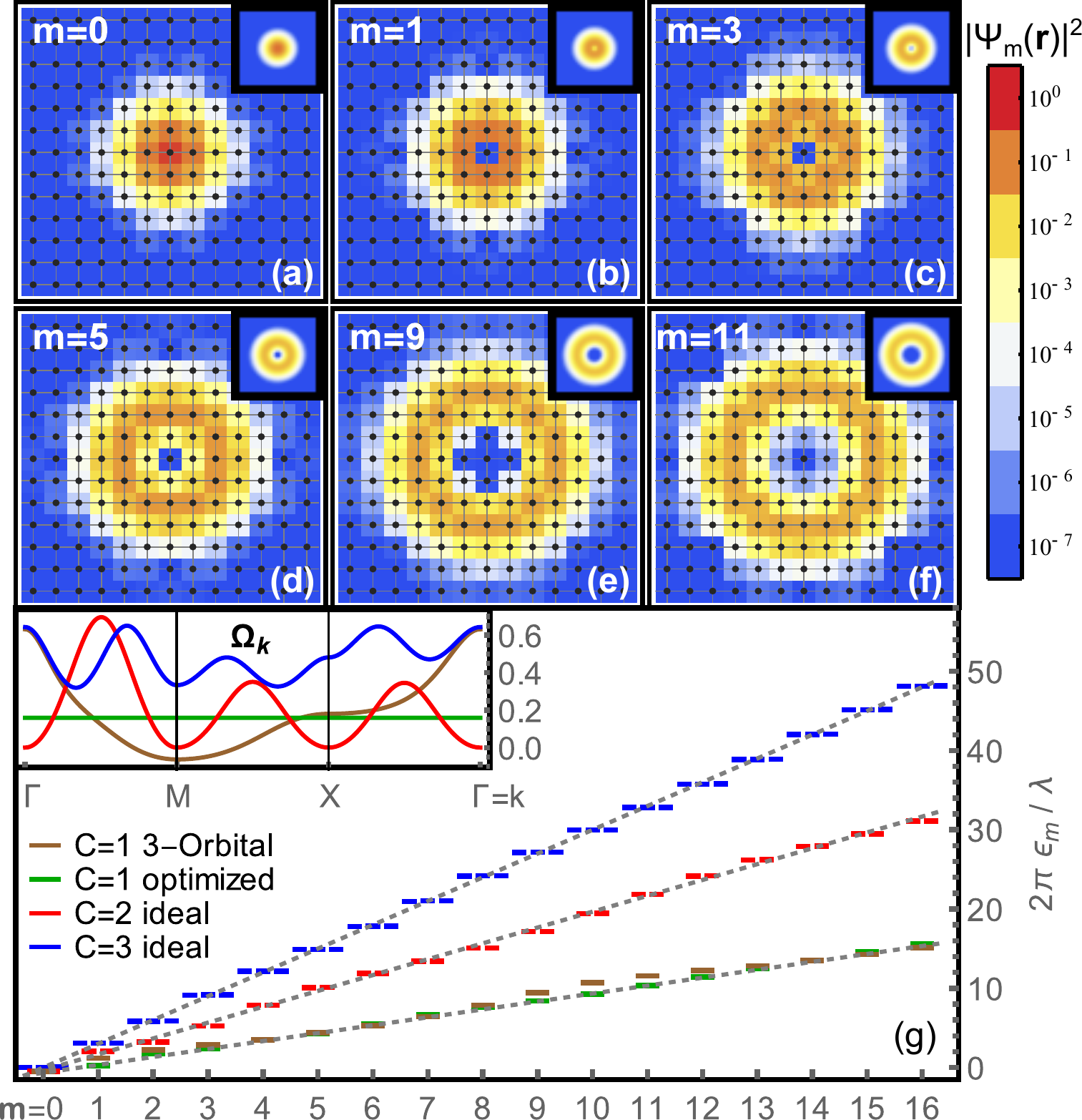}
	\caption{(color online). (a-f) Real-space lattice guiding-center wave functions for the ideal isotropic $\Ch=3$ model, evaluated from (\ref{eq:HamConfinementAnisotropic}). Insets depict analogous conventional LLL wave functions.
(g) Berry curvature $\Omega_\k$ (inset) and associated MLL spectra for the FCI models discussed in the main text. $m$ is the guiding center index; dotted lines indicate ideal flat-$\Omega$ spectrum $\E_m = \frac{\lambda \Ch m}{2\pi}$. The $\Ch$-fold degeneracy of MLLs reflects the $\Ch$-component basis for $\Ch>1$ models. The guiding-center structure remains robust in the presence of fluctuations of $\Omega_\k$.} \label{fig:singlebodyPhysics}
\end{figure}

\paragraph{Isotropic case:} Physical insight may be gleaned by identifying (\ref{eq:confinementHamiltonian0}) with an electron in a magnetic field but in momentum-space; $\eta^{\mu\nu} = \delta^{\mu\nu}$ is particularly instructive:
\begin{align}
	&\HamConf = \frac{\lambda}{2} \left[ -i \boldsymbol{\nabla}_{\k} + \BerryConn(\k) \right]^2 + \frac{\lambda}{2} \tr \mathbf{g}(\k) \hspace{0.7cm} \k \in {\rm BZ}
\end{align}
Here, $\Omega(\k)$ identifies with the magnetic field, and electrons scattering at small momenta pick up Berry phase factors in analogy to the Aharonov-Bohm effect. 
For benign 'magnetic field' fluctuations over the BZ $\int d^2k [k_B^2 \Omega(\k) - 1]^2/A_{BZ} < 1$ with inverse magnetic length $k_B^2 = \frac{A_{BZ}}{2\pi\mathcal{C}}$ and BZ area $A_{BZ}$, the guiding-center basis is therefore described by the well-known Landau levels on the torus penetrated by flux $\Ch$, but \textit{in momentum-space}. These momentum-space Landau levels (MLLs) are indexed by two quantum numbers $m,n$ with eigenspectrum $\E_{mn} = \frac{\lambda \Ch}{\sqrt{A_{BZ}}} (m+1)$, where $m$ is the MLL index and $n$ indexes a $\Ch$-fold degeneracy per MLL \cite{HaldaneRezayi1985}. Identification of the usual guiding center coordinates is thus reversed: the MLL index $m$ plays the role of the FCI guiding center index and can be identified with discrete $C_N$ rotational symmetry if present, whereas $n = 0,...,\Ch-1$ acts as a component index for $\Ch>1$ \cite{footnoteConjugateOscillatorLandauLevel}. In real space, MLL wave functions are radially-localized (Fig. \ref{fig:singlebodyPhysics}(a-f)). Importantly, (\ref{eq:confinementHamiltonian0}) does not enter as physical confinement in the infinite system, thus $\lambda \to 0$ can be taken in the thermodynamic limit. However, $\lambda\neq 0$ enters as a proper energy scale when considering finite-size droplets \cite{footnoteTruncation, ScaffidiPRB2014}. 

\paragraph{Anisotropic case:} Generically, the confinement metric $\eta^{\mu\nu}$ can be expressed in terms of a complex vector $\boldsymbol{\omega}$ that obeys $\eta^{\mu\nu} = \bar{\omega}^\mu \omega^\nu + \bar{\omega}^\nu \omega^\mu$. The isotropic limit becomes $\bar{\omega} = [ 1,i ]^\top/\sqrt{2}$. Corresponding guiding center operators
\begin{align}
	&\PI = \PII_\mu~ \omega^\mu ~,~~~\PID = \bar{\omega}^\mu ~ \PII_\mu    \label{eq:guidingcenterOps}
\end{align}
obey commutation relations $[\PI,\PID] = \Omega$ when $\partial_{k_\mu}\omega^\mu=0$, which fixes a phase freedom $\boldsymbol{\omega} \to \boldsymbol{\omega}e^{i\varphi}$.
Substituting (\ref{eq:guidingcenterOps}) in (\ref{eq:confinementHamiltonian0}) yields the confinement Hamiltonian in guiding-center language that determines the single-body basis:
\begin{align}
	\HamConf = \lambda~ \PI \PID + \tfrac{1}{2} \lambda \left( \eta^{\mu\nu} g_{\mu\nu} - \Omega \right)       \label{eq:HamConfinementAnisotropic}
\end{align}
Comparison with Haldane's construction \cite{HaldanePRL2011} reveals that $\eta^{\mu\nu}$ is precisely the FCI momentum-dual of the \textit{guiding-center metric} of the anisotropic FQHE. The root cause for this duality can be inferred by noting that the conventional FQHE can in fact be formulated \textit{both} in position and momentum representation, with single-body dynamics and LLL wave functions form-invariant under interchange of complex coordinates $x+iy$ and momenta $k_x+ik_y$. This situation is drastically different in FCIs, where the discreteness of the lattice necessitates switching to momentum space in order to retain a first-quantized description in terms of continuous coordinates.

Conceptually, the challenge of devising a microscopic description of FQH states in FCIs reduces to a variational problem of determining the deformation of the guiding-center orbitals upon placing a FQH liquid on the lattice. Given an appropriate choice of $\eta^{\mu\nu}(\k)$, \textit{any FCI} can in principle be captured by many-body trial ground states, constructed from the single-body eigenstates of (\ref{eq:HamConfinementAnisotropic}): for instance, given the lowest MLL wave function $\Psi_0(\k)$ and ladder operators $\AD{}$ that generate higher MLLs, the Laughlin state at $\nu$ reads $\Psi_{\nu} \sim \prod_{i<j} ( \AD{i} - \AD{j} )^{1/\nu} \Psi_0(\k)$.

A \textit{preferred} guiding-center metric can be readily identified by demanding suppression of a residual dispersive term $\eta^{\mu\nu} g_{\mu\nu} - \Omega$ in (\ref{eq:HamConfinementAnisotropic}) that delocalizes the MLL basis:
\begin{align}
	\boldsymbol{\eta}(\k) = \sqrt{\det \mathbf{g}(\k)}~ \mathbf{g}^{-1}(\k)      \label{eq:confinementMetricFixing}
\end{align}
The dispersion vanishes exactly if and only if
\begin{align}
	2\sqrt{\det \mathbf{g}(\k)} = |\Omega(\k)|   \label{eq:IdealDroplet}
\end{align}
This is the \textit{condition for an ideal FCI droplet}, and is satisfied by every two-band model \cite{SUPPLEMENTARYdetailedderivation} while placing constraints on models with three or more bands.

\paragraph{Interactions:} Insight into stabilization of trial states necessitates recasting the many-body problem into first quantization. Analogous to the FQHE, density interactions may be generalized via an \textit{interaction metric} $\tilde{\eta}^{\mu\nu}$:
\begin{align}
	\Ham_I = \sum_\q V_\q~ \dens_\q \dens_{-\q} ,~~~ \dens_\q = \sum_{\alpha\k} \CD{\k\alpha} \TOp_\q[\tilde{\eta}^{\mu\nu}]~  \C{\k\alpha}      \label{eq:HamInteractionBare}
\end{align}
where $\TOp_\q[\tilde{\eta}^{\mu\nu}] = e^{q_\mu \tilde{\eta}^{\mu\nu} \partial_{k_\nu}}$ enters as a metric-dependent momentum-space translation operator. In contrast to the FQHE on the plane, translation symmetry and periodicity in $\q$ constrain $\tilde{\eta}^{\mu\nu}$ to ${\rm SL}(2,\mathbb{Z})$. Importantly, the guiding center metric is still a variational degree of freedom if the deviation of (\ref{eq:confinementMetricFixing}) and $\tilde{\eta}^{\mu\nu}$ is significant.
To proceed, note that if the ideal droplet condition (\ref{eq:IdealDroplet}) is satisfied, then an exact operator identity $\left[ 1 - \ketbra{u_\k}{u_\k} \right] \bar{\omega}^\mu \partial_{k_\mu} \ketbra{u_\k}{u_\k} = 0$ \cite{RoyARXIV2012} entails that any operator of the form $\hat{\mathcal{O}} = \hat{\Lambda}^{-}(-i\partial_\mu \omega^\mu) \hat{\Lambda}^{+}(-i \bar{\omega}^\mu \partial_\mu)$ with analytic functions $\Lambda^\pm$ can be projected to the flat band as $\hat{P} \hat{\mathcal{O}} \hat{P}^\top  = \hat{\Lambda}^{-}(\PI) \hat{\Lambda}^{+}(\PID)$. Consider thus a decomposition of $\tilde{\eta}^{\mu\nu} = \bar{\chi}^\mu \omega^\nu + \chi^\mu \bar{\omega}^\nu$ with complex vector $\boldsymbol{\chi}$: if $\boldsymbol{\chi}$ is momentum-independent, then the translation operator is precisely of this form: $\TOp_\q = e^{q_+ \omega^\mu \partial_{k_\mu}} e^{q_- \bar{\omega}^\mu \partial_{k_\mu}}$ with $q_{\pm} = q_\mu \bar{\chi}^\mu, q_\mu \chi^\mu$ --- this case is considered in detail below \cite{footnoteGeneralTranslationOperator}. Note that in the isotropic limit $\tilde{\eta}^{\mu\nu}, \eta^{\mu\nu} = \delta^{\mu\nu}$, $\TOp_\q$ reduces to the conventional translation operator $\hat{\mathcal{T}}_\q = e^{i\q\cdot(-i\boldsymbol{\nabla}_\k)}$ with $q_{\pm} = (q_x \pm i q_y)/\sqrt{2}$, while $H_I$ is just the usual density interaction. While emphasis has thus far been placed on narrowing down to a suitable class of interactions, substantial progress has been made: projected to the flat band, the joint dynamics of (\ref{eq:ConfinementPotential}), (\ref{eq:HamInteractionBare}) can now be succinctly expressed in guiding-center language:
\begin{align}
	\Ham = \lambda \sum_i \PI_i \PID_i + \sum_{i<j \q} V_\q e^{i q_+ (\PI_i - \PI_j)} e^{i q_- (\PID_i - \PID_j)}      \label{eq:HamFull}
\end{align}

This Hamiltonian is the central result of this paper - it provides a first-quantized description of the low-energy dynamics of an ideal FCI in terms of the quantum geometry $\eta^{\mu\nu}, \tilde{\eta}^{\mu\nu}$ of the lattice. Its interaction describes two-body momentum-space magnetic translations, and acts solely on the relative guiding center indices.
A key consequence is the approximate conservation of center-of-mass guiding center, quantified by Berry curvature fluctuations averaged over the BZ with
$\frac{A_{\BZ}}{2} \| [ \PI_{\rm rel} , \PID_{\rm cm} ] \|^2 = \int_\BZ d^2\k~ [ \Omega(\k) - \frac{2\pi \Ch}{A_{\BZ}} ]^2$,
where $\PI_{\rm cm/rel} = (\PI_1 \pm \PI_2)/\sqrt{2}$ span the two-body problem in the Chern band \cite{footnoteTwoBodyProblem}. Furthermore (\ref{eq:HamFull}) does not act on the intra-MLL index $n$, stipulating an emergent SU($\Ch$) symmetry for $\Ch>1$ \cite{footnoteSUCsymmetry}.

The physics of the above Hamiltonian can be studied via a pseudopotential decomposition of the two-body interaction matrix elements $V_{mm'}^{MM'}$ $= \left<mM\right|\Ham\left|m'M'\right>|_{\lambda= 0}$ with $m,M$ relative and center-of-mass guiding center indices, and intra-MLL indices omitted. Approximate center-of-mass conservation in (\ref{eq:HamFull}) entails that the two-body repulsion depends only on the relative coordinate, $V_{mm'}^{MM'} \approx  V_{mm'} \delta_{MM'}$. Since the guiding center index identifies with discrete rotational symmetries, it is tempting to speak of an emergent continuous rotational symmetry in the flat band -- however, it persists even for $V_\q$ anisotropic; $C_N$ symmetry of $V_\q$ instead constrains relative guiding center transitions $V_{m\neq m'}$ to $(m-m') {\rm mod} N = 0$. The dominant matrix elements are thus well-captured by Haldane pseudopotentials \cite{HaldanePseudopotentials}  $V_m = \frac{1}{\mathcal{M}} \sum_{M=0}^\mathcal{M} V_{mm}^{MM} |_{\mathcal{M}\to\infty}$, which indicate stabilization of FQH trial states.

The propensity of the MLL basis to lead to a well-defined pseudopotential expansion for such models is a key advantage of the first-quantized formalism.
Treating $\Omega(\k)$ fluctuations as a perturbation with ladder operators $\PI_{i} \rightarrow \A{i}$, $\ket{m,M} = (\AD{\rm rel})^m (\AD{\rm cm})^M \ket{0,0}/\sqrt{m!M!}$ leads to
\begin{align}
	V_{mm'} = \int d\q \frac{ V_\q (iq_-)^{m-m'}  \Kummer{m+1}{m-m'+1}{-\frac{2q_+ q_-}{k_B^2}} }{k_B^{m-m'}\sqrt{2^{m'-m} m'! / m!}(m-m')!}   \label{eq:pseudopotentials}
\end{align}
Here, ${}_1F_1(\cdot)$ is the Kummer confluent hypergeometric function. The well-known pseudopotentials of the conventional FQHE can be readily recovered for $m=m'$, $q_\pm = (q_x\pm iq_y)/\sqrt{2}$ with $V_m = \int d^2\q~ V_\q L_m(\frac{\q^2}{\Omega}) e^{-\q^2/\Omega}$.

\begin{figure}
	\includegraphics[width=8.9cm]{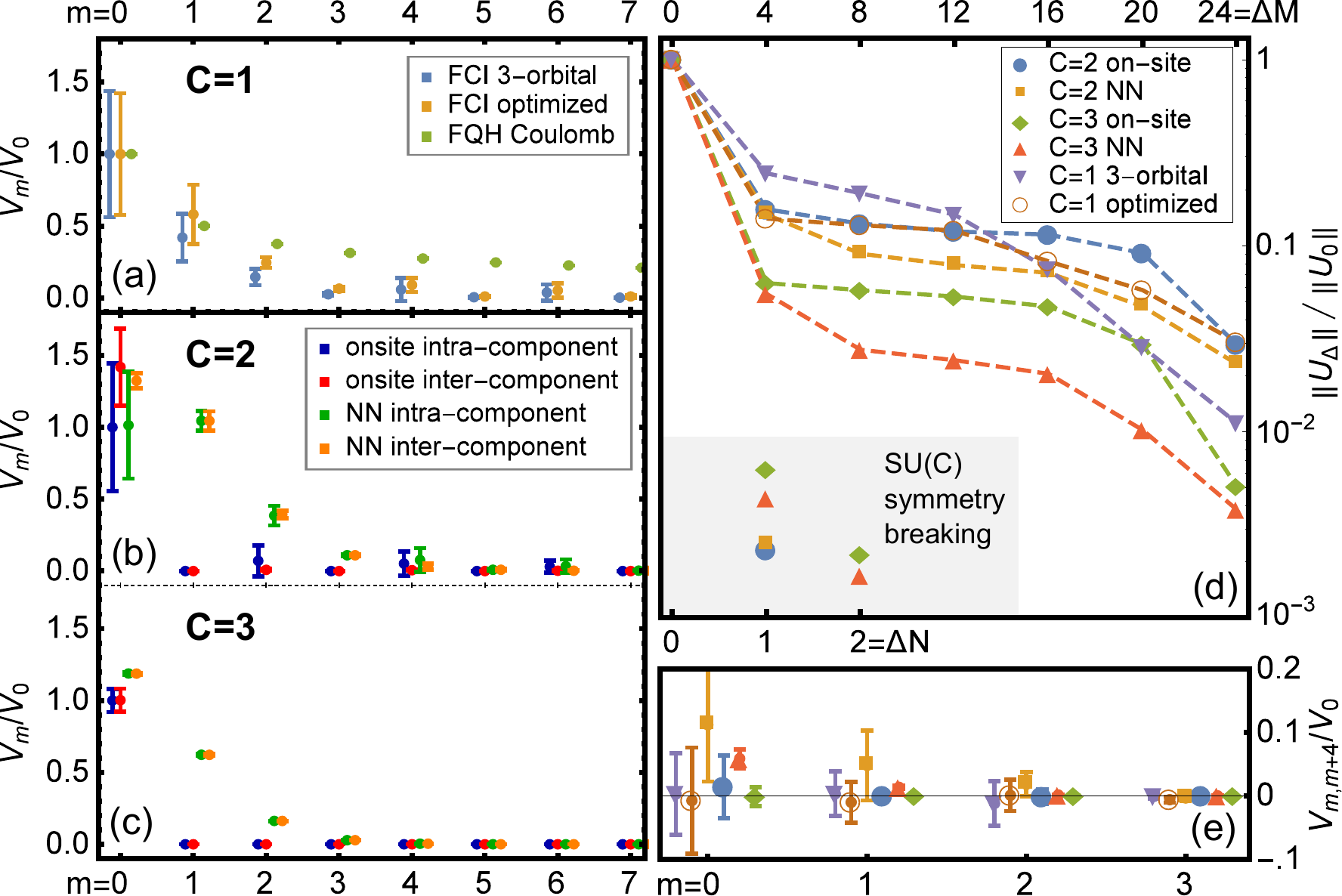}
\caption{(color online). (a)-(c) Haldane pseudopotentials for the $\Ch=1$ three-orbital and $\Ch>1$ ideal droplet models (on-site and NN interactions). Only odd (even) pseudopotentials determine interactions between same-species fermions (bosons). Error bars quantify the residual center-of-mass $M$ variation, strongly suppressed for the ideal $\Ch=2,3$ models due to emergent conservation of guiding center.
(d) Log plot of guiding-center and $SU(\Ch)$ symmetry breaking in the interacting problem with matrix elements $V_{m_1\dots m_4}^{n_1\dots n_4}$, quantified as the ratio of the norm of $\Delta M = m_1+m_2-m_3-m_4$ ($\Delta N = n_1 + n_2 - n_3 - n_4 ~{\rm mod}~ \mathcal{C}$ component) symmetry-breaking and -preserving two-body matrix elements. The guiding-center index identifies with C$_4$ symmetry.
(e) Role of broken rotational symmetry on the lattice, depicted via leading-order $V'_{m,m+4}$ of generalized pseudopotentials $V'_{mm'} = \bra{m,M}H\ket{m',M}$. 
}
\label{fig:interactions}
\end{figure}

\paragraph{Isotropic ideal FCI models:} While the focus so far has been placed on developing an accurate language for the generic anisotropic case, a key follow-up question concerns instead applying above results to find an FCI analog of the \textit{isotropic} Landau level, particularly favorable for FQH phases.  Such models indeed exist: the isotropic case $\eta^{\mu\nu},\tilde{\eta}^{\mu\nu} = \delta^{\mu\nu}$ with corresponding Fubini-Study metric $g_{xx} - g_{yy}, g_{xy} = 0, \tr \mathbf{g}(\k) = \Omega(\k)$ is \textit{uniquely satisfied} by any Bloch state that can be written without normalization as a \textit{meromorphic function} $\ket{\bar{u}_\k} = \ket{\bar{u}_{k_x + i k_y}}$. The number of poles in the BZ defines $\Ch$ \cite{ChaomingPSS}; periodic boundary conditions in $\k$ restricts $\ket{\bar{u}_{k_x + i k_y}}$ to elliptic functions, constrained to $\Ch \geq 2$. Skew-anisotropic guiding center metrics ensue from distortions of the lattice: for instance, a 'squeezed' FQH liquid with $\eta_{xx} = 1/\zeta, \eta_{yy} = \zeta$ follows from BZ strain deformations with $k_x \to \zeta k_x$.

To illustrate our construction, we consider two multi-orbital toy models on the square lattice for $\Ch=2,3$ \cite{FORTHCOMING}, with the guiding-center basis of MLLs depicted in fig. \ref{fig:singlebodyPhysics}:
\begin{align}
	\ket{\bar{u}_\k^{\Ch=2}} &= \left[\begin{array}{ll} 1, & \alpha~ \wp(k_x+ik_y) \end{array}\right]^\top      \label{eq:C2Model}   \\
	\ket{\bar{u}_\k^{\Ch=3}} &= \left[\begin{array}{lll} 1, & \beta~ \wp(k_x+ik_y), & \gamma~ \wp'(k_x+ik_y) \end{array}\right]^\top     \label{eq:C3Model}  
\end{align}
Here, $\wp(z)$ is the Weierstrass elliptic function with periods $2\pi$, $2\pi i$ and a second-order pole at the $\Gamma$ point, and $\alpha=5.77$, $\beta=-7.64$, $\gamma=6.73$ are band structure parameters, chosen to minimize Berry curvature fluctuations. 

The corresponding Bloch Hamiltonian is not unique; a possible definition with flat bands is $\mathbf{h}_{\Ch}(\k) =$ $\mathbf{1} - \ketbra{\bar{u}_k^{\Ch}} {\bar{u}_k^{\Ch}} / {\braket{\bar{u}_k^{\Ch}}{\bar{u}_k^{\Ch}}}$. In general, $\mathbf{h}_{\Ch}(\k)$ exhibits long-ranged but exponentially-decaying hopping terms \cite{ChaomingPSS} and acts as an artificial toy lattice model for FCI states with topological order, much like the Hubbard model truncated to nearest-neighbor (NN) hoppings acts as a toy model for strongly-correlated states with conventional order. While a classification of such ideal FCI host lattice models remains an important open task, note that simple physical models can emerge after truncation of irrelevant hoppings. For instance, $\mathbf{h}_{\Ch=2}$ is well-described by a canonical d-wave lattice model $\mathbf{h}_{\Ch=2}(\k) = \mathbf{d}(\k)\cdot \vec{\boldsymbol{\sigma}}$ with $\mathbf{d}(\k) = \left[ t(\cos k_x + \cos k_y ),~ t(\cos k_x - \cos k_y ),~ t' \sin(k_x)\sin(k_y) \right]^\top$ with nearest- and next-nearest-neighbor hoppings.

\paragraph{Analysis:} 
In addition to the ideal isotropic models (\ref{eq:C2Model},\ref{eq:C3Model}), we study both a conventional three-orbital $\Ch=1$ model on the square lattice \cite{SunPRL2011} that does not satisfy (\ref{eq:IdealDroplet}), and an ``optimized'' variant with flat Berry curvature (${\rm var}(\Omega_\k) \approx 10^{-6}$), obtained via adiabatically adding symmetry-preserving hoppings up to 5$^{\rm th}$ neighbor \cite{SUPPLEMENTARYdetailedderivation}.

Fig. \ref{fig:interactions} depicts the pseudopotential decomposition for on-site and NN repulsion, using the MLL basis of Fig. \ref{fig:singlebodyPhysics}.
The $\Ch=2,3$ ideal lattice models display emergent guiding-center and SU($\Ch$) symmetries, manifested in a vanishing center-of-mass dependence of pseudopotentials (figs. \ref{fig:interactions}b, \ref{fig:interactions}c), and suppression of symmetry-breaking two-body interaction terms (fig. \ref{fig:interactions}d).  As anticipated from (\ref{eq:pseudopotentials}), on-site repulsion results in a non-zero pseudopotential only for $V_0$, acting as an optimal (221)-Halperin state \cite{Halperin1983} parent Hamiltonian for hardcore bosons, whilst not stabilizing fermionic FQH states \cite{footnoteFermionpseudopotentials}. The latter can be remedied by tuning NN repulsion to tune $V_1$, $V_3$. The controlled expansion in pseudopotentials is a key merit of this construction and highlights that, contrary to common perception, the confluence of flat Berry curvature and local interaction \textit{does not} stabilize a fermionic FQH liquid in an FCI.
Conversely, the non-optimal $\Ch=1$ models violate (\ref{eq:IdealDroplet}) and do not pin the guiding-center metric to $\mathbf{g}$. The guiding-center description (\ref{eq:HamFull}) is incomplete, broadening the effective interaction range with non-vanishing and decaying $V_m$ with substantial center-of-mass deviation even for a purely local interaction (fig. \ref{fig:interactions}a) that persists even for uniform Berry curvature.
We stress that this highlights the shortcomings of long-wavelength limit arguments \cite{ParameswaranPRB2012, EstiennePRB2012, RoyARXIV2012, GoerbigEPJ2012, MurthyShankarPRB2012,DobardzicPRB2013, RepellinPRB2014} in predicting the microscopics of the FQHE on the lattice.

In summary, we introduced a first-quantized description of FCIs, with the FQHE emerging in a picture of anisotropic momentum-space Landau levels in a fluctuating magnetic field. We presented a novel class of ideal FCI lattice models as duals of the isotropic FQHE and demonstrated their optimality via an expansion of local interactions into Haldane pseudopotentials which can be determined straightforwardly in first quantization. A primary goal of this work is to establish a deeper microscopic understanding of the stabilization of FQH states in flat Chern bands - the resulting interplay of topology and geometry to determine long- and short-wavelength physics on the lattice serves as a natural application of the formalism of the anisotropic FQHE. The results presented set a foundation for microscopic analysis of non-Abelian phases on the lattice and extension to fractional topological insulators.

\acknowledgements{
We acknowledge support from the U. S. Department of Energy, Office of Basic Energy Science, Division of Materials Science and Engineering under Contract No. DE-AC02-76SF00515. C.H.L. is supported by a fellowship from the Agency of Science, Technology and Research of Singapore. R.T. is supported by the European Research Council through ERC-StG-336012-TOPOLECTRICS. X.L.Q. is supported by the David \& Lucile Packard Foundation.
}

\pagebreak
\onecolumngrid
\appendix
\pagebreak

\begin{center}
\textbf{\large Supplementary Material}
\end{center}

\section{Momentum-Space Landau Levels}

As described in the main text, the guiding-center basis in a FCI can be understood in terms of momentum-space Landau levels (MLLs).
We seek the low-energy momentum-space dynamics induced by a real-space confinement potential $\hat{V}(\r) = \frac{\lambda}{2} x_{\mu} \eta^{\mu\nu} x_{\nu}$.
The confinement metric $\eta^{\mu\nu}$ can be expressed as an $N_d$-fold superposition of reciprocal lattice vectors $\eta^{\mu\nu} = \sum_i^{N_d} \eta_i d_i^{\mu} d_i^{\nu}$, where $\mathbf{d}_i = m_1 \mathbf{b}_1 + m_2 \mathbf{b}_2$ and $m_1, m_2 \in \mathbb{Z}$.
A minimum of three basis vectors $\mathbf{d}_i$ is required to reproduce the three independent components of $\boldsymbol{\eta}$.
Regularization of the long-distance behavior of the confinement potential on an $L \times L$ lattice is not unique; a simple choice reads:
\begin{align}
	\hat{V}_{\rm reg} = \frac{\lambda}{2} \frac{L^2}{(2\pi)^2} \sum_{i}^{N_d} \eta_i \left[ 1 - \cos( \mathbf{d}_i \cdot \mathbf{r} / L ) \right]
\end{align}
The regularized form of the confinement potential is used for numerical results presented in the main text. Choosing a sufficiently large lattice, one can solve for the guiding-center basis as eigenstates of $\tilde{H}_0 + \hat{V}_{\rm reg}$, where care must be taken to ensure that the band gap of $\tilde{H}_0$ is much larger than the depth of the confinement potential. The lattice dimensions are taken to be 160x160 sites for numerical results in this work. As is canonical in numerical studies of FCIs, we assume that the energy scale of interactions in the flat band is much greater than its band width and work with a modified lattice Hamiltonian $\tilde{H}_0 = \frac{1}{\E_\k} \Ham_0$ to eliminate residual dispersion $\E_\k$ in the flat band. This step is not necessary for the ideal models of equations (14), (15) in the main text, as they intrinsically exhibit a perfectly flat Chern band. The low-energy dynamics follow from projecting $\hat{V}_{\rm reg}$ onto the reduced Hilbert space built from the set of Bloch states $\ket{u_\k}$ spanning the flat band, and taking the thermodynamic limit:
\begin{align}
	\hat{H} &= \lim_{L\to\infty}  \frac{\lambda}{2} \sum_{\k\k' \in\BZ} \ketbra{\k}{u_\k} \hat{V_{\rm reg}} \ketbra{u_{\k'}}{\k'} \notag\\
	&= \lim_{L\to\infty}  \frac{\lambda L^2}{4} \sum_{\k\in\BZ} \ketbra{\k}{u_\k} \sum_i^{N_d} \eta_i \left[ 2 \ketbra{u_\k}{\k} -  \ketbra{u_{\k+\frac{\mathbf{d}_i}{L}}}{\k+\frac{\mathbf{d}_i}{L}} - \ketbra{u_{\k-\frac{\mathbf{d}_i}{L}}}{\k-\frac{\mathbf{d}_i}{L}} \right] \notag\\
	&= \frac{\lambda}{2} \sum_i^{N_d} \eta_i \int_{\BZ} d^2\k~ \ketbra{\k}{u_\k} \left( -i \mathbf{d}_i \cdot \nablaK \right)^2 \ketbra{u_\k}{\k} \notag\\
	&= \frac{\lambda}{2} \int d^2\k~ \left(\sum_i^{N_d} \eta_i d_i^{\mu} d_i^{\nu}\right) \ket{\k} \left[\vphantom{\frac{1}{2}} (-i\partial_\mu)(-i\partial_\nu) + i A_\mu \partial_\nu + i A_\nu \partial_\mu - \left( \braOPket{u_\k}{\partial_\mu \partial_\nu}{u_\k} \right) \right] \bra{\k} \notag\\
%
%	&= \frac{\lambda}{2} \int d^2\k~ \ket{\k} \eta^{\mu\nu} \left[ \left( -i\partial_\mu + A_\mu \right) \left( -i\partial_\nu + A_\nu \right) - \left( \braOPket{u_\k}{\partial_\mu \partial_\nu}{u_\k} \right) - A_\mu A_\nu + i\left(\partial_\mu A_\nu\right) \right] \bra{\k} \notag\\
%
	&= \frac{\lambda}{2} \int d^2\k~ \ket{\k} \left[  \left( -i\partial_\mu + A_\mu \right) \eta^{\mu\nu} \left( -i\partial_\nu + A_\nu \right) + \eta^{\mu\nu} g_{\mu\nu} \right] \bra{\k} \label{eq:SUPPeffHam}
\end{align}
where $A_\mu = -i\braOPket{u_\k}{\partial_\mu}{u_\k}$ is the Berry connection, $\Omega = \epsilon^{\mu\nu} \partial_\mu A_\nu$ is the Berry curvature, and the Fubini-Study metric $g_{\mu\nu}$ is defined as:
\begin{align}
	g_{\mu\nu} &= \frac{1}{2} \left[\spacer \braket{\partial_\mu u_\k}{\partial_\nu u_\k} - \braket{\partial_\mu u_\k}{u_\k} \braket{u_\k}{\partial_\nu u_\k} + \left( \mu \leftrightarrow \nu \right) \right]
\end{align}
The last line of equation \ref{eq:SUPPeffHam} uses the fact that the metric tensor $\eta^{\mu\nu}$ is symmetric, hence $\eta^{\mu\nu} \epsilon_{\mu\nu} = 0$. In terms of operators $\PII_\mu = -i\partial_\mu + A_\mu$, one finally arrives at the first-quantized Hamiltonian (4) described in the main text.

A smooth gauge choice separates $\BerryConn(\k) = \frac{\pi\mathcal{C}}{A_{BZ}} [ -k_y, ~ k_x ]^\top + \BerryConn_{\rm fluct}(\k)$ into a topological contribution and fluctuations. $\Ch \neq 0$ then stipulates that wave functions acquire twisted boundary conditions $\psi_{nm}(\k+m_1\mathbf{b}_1+m_2\mathbf{b}_2) = e^{-i\frac{\Ch}{4\pi} (m_1 \mathbf{a}_2 - m_2 \mathbf{a}_1)\k} \psi_{nm}(\k)$ that constrain the intra-MLL index to $0\dots\Ch-1$, giving rise to the MLL spectrum described in the main text. As a side note, residual dispersion in the Chern band can be understood as a momentum-space confinement potential, giving rise to a Darwin-Fock eigenbasis \cite{DarwinFock} on the torus, with the ratio of confinement $\lambda$ and bandwidth entering as a variational degree of freedom when building the single-body basis.

\section{Ideal Droplet Condition -- Two-Band Models}

The ideal droplet condition $2 \sqrt{\det \mathbf{g}(\k)} = \Omega(\k)$
is automatically satisfied by any two-band FCI model. To see this, consider a generic two-band Bloch Hamiltonian $\mathbf{h}(\k) = \E(\k)~\hat{\mathbf{d}}(\k) \cdot \vec{\boldsymbol{\sigma}}$
parameterized by unit vector $\hat{\mathbf{d}}(\k)$ and dispersion $\pm \E(\k)$, where $\vec{\boldsymbol{\sigma}}$ are the Pauli matrices in orbital basis. The Berry curvature and Fubini-Study metric read:
\begin{align}
	\Omega(\k) &= \frac{1}{2} \hat{\mathbf{d}}(\k) \cdot \frac{\partial \hat{\mathbf{d}}(\k)}{\partial k_x} \times \frac{\partial \hat{\mathbf{d}}(\k)}{\partial k_y} \\
	g_{\mu\nu}(\k) &= \frac{1}{4} \left(\frac{\partial \hat{\mathbf{d}}(\k)}{\partial k_\mu} \times \hat{\mathbf{d}}(\k)\right) \cdot \left(\hat{\mathbf{d}}(\k) \times \frac{\partial \hat{\mathbf{d}}(\k)}{\partial k_\nu} \right)
\end{align}
The determinant of the Fubini-Study metric follows from application of quadruple-product identities and satisfies $2 \sqrt{\det \mathbf{g}(\k)} = \Omega(\k)$.

\section{Haldane Pseudopotentials}

This section discusses the derivation of Haldane pseudopotentials for momentum-space Landau levels. The starting point is the two-body interaction in first quantization that follows from equation (12) in the main text.
%\begin{align}
%	\Ham_I = \sum_{\q} V_\q e^{i q_+ (\PI_1 - \PI_2)} e^{i q_- (\PID_1 - \PID_2)}
%\end{align}
The guiding-center basis can be found via numerical solution of the single-body problem. Given the lowest MLL state $\ket{\boldsymbol{\Psi}_0}$ and a set of ladder operators $\A{}$, $\AD{}$, $\comm{\A{i}}{\AD{i}} = 1$ generating the $m$-th MLL $\ket{m} = \frac{(\AD{})^m}{\sqrt{m!}} \ket{\boldsymbol{\Psi}_0}$, one can define two-body states
\begin{align}
	\ket{m,M} &= \frac{1}{\sqrt{m!M!}} \left(\frac{\AD{1}-\AD{2}}{\sqrt{2}}\right)^m \left(\frac{\AD{1}+\AD{2}}{\sqrt{2}}\right)^M \ket{\boldsymbol{\Psi}_0}
\end{align}
The Haldane pseudopotentials follow
\begin{align}
	V_{mM} &= \braOPket{m,M}{ \sum_{\q} V_\q e^{i q_{+} (\PI_1 - \PI_2)} e^{i q_{-} (\PID_1 - \PID_2) }}{m,M}
\end{align}
and can be evaluated numerically, with the residual center-of-mass dependence studied in the main text. To recover the limit of flat Berry curvature, note that the guiding center operators can be written in terms of the ladder operators $\PI_i = \A{i} / k_B$ with $k_B = \sqrt{2\pi/\mathcal{C}}$ the magnetic wave vector for a square lattice, which allows for an algebraic evaluation of the two-body matrix elements:
\begin{align}
	V_{\substack{mM \\ m'M'}} &= \int d^2q~ V_q~ \braOPket{mM}{\hat{V}}{m'M'} \notag\\
%	&= \int d^2q~ V_q~\frac{1}{\sqrt{m!M!m'!M'!}\sqrt{2^{m+m'+M+M'}}} ~\times \notag\\
%	&~~~~~~~\times~\braOPket{0}{\left(\A{1}+\A{2}\right)^M \left(\A{1}-\A{2}\right)^m e^{i q_+ (\A{1}-\A{2})/k_B} e^{i q_- (\AD{1}-\AD{2})/k_B} \left(\AD{1}-\AD{2}\right)^{m'} \left(\AD{1}+\AD{2}\right)^{M'} }{0} \notag\\
	&= \int d^2q~ V_q~ \frac{1}{\sqrt{m!m'!}} \braOPket{0}{ \left(\A{1}-\A{2}\right)^m e^{i q_+ (\A{1}-\A{2})/k_B} e^{i q_- (\AD{1}-\AD{2})/k_B} \left(\AD{1}-\AD{2}\right)^{m'} }{0}  \delta_{MM'} \notag\\
	&= \int d^2q~ V_q~ \frac{1}{\sqrt{m!m'!}} \sum_{ll'} \frac{(\sqrt{2} i q_+/k_B)^l (\sqrt{2} i q_-/k_B)^{l'}}{l!l'!} \braOPket{0}{ \left(\A{}\right)^{m+l}  \left(\AD{}\right)^{m'+l'} }{0} \delta_{MM'}  \notag\\
	&= \int d^2q~ V_q~ \frac{1}{\sqrt{m!m'!}} \sum_{l} \frac{(\sqrt{2} i q_+/k_B)^l (\sqrt{2} i q_-/k_B)^{m-m'+l}}{l!(m-m'+l)!} (m+l)! \delta_{MM'}  \notag\\
	&= \int d^2q~ V_q~ \sqrt{\frac{m!}{m'!}}\frac{(\sqrt{2}i q_-/k_B)^{m-m'}~  {}_1F_{1}\left(m+1;m-m'+1,-\frac{2\bar{q}^2}{k_B^2}\right) }{(m-m')!} \delta_{MM'} 
\end{align}
where ${}_1F_{1}(a;b;z)$ is the Kummer confluent hypergeometric function. Here $q_+ = \bar{\chi}^\mu q_\mu$, $q_- = \chi^\mu q_\mu$, and $\bar{q} = q_- q_+ = q_\mu \bar{\chi}^\mu \chi^\nu q_\nu$. While the center of mass guiding center dependence is eliminated, lack of continuous rotational symmetry on the lattice does not forbid subdominant scattering between states with different relative guiding center indices $m,m'$ unless the parent interaction is purely local $V_\q \to V$. Instead, $C_N$ symmetry can provide weaker selection rules with $V_{mm'} \neq 0$ only for $m - m' ~{\rm mod}~ N = 0$. The conventional Haldane pseudopotentials follow from imposing $m=m'$, $q_{\pm} = (q_x \pm i q_y)/\sqrt{2}$, and expressing the hypergeometric function in terms of a Laguerre polynomial:
\begin{align}
	V_m &= \int d^2\q~ V_{\q} L_m\left(\frac{\q^2}{k_B^2}\right) e^{-\q^2/k_B^2}
\end{align}
which is the formula quoted in the main text.

\section{Elliptic Function Models}

The main text describes a class of ideal FCI models which can be defined from unnormalized Bloch states $\ket{\tilde{u}_{k_x+i k_y}}$ written as elliptic functions. To see this, consider a generic N-band model with unnormalized Bloch state written in terms of complex momenta
\begin{align}
	\ket{\tilde{u}_{\k}} = \left[ \phi_1(z,\bar{z}),~\phi_2(z,\bar{z}),\dots,\phi_N(z,\bar{z}) \right]^\top
\end{align}
where $z,\bar{z} = k_x \pm i k_y$. The corresponding Berry curvature and Fubini-Study metric may be evaluated straight-forwardly:
\begin{align}
	\Omega &= 2\sum_{n>m}^N \left[ \frac{\left|\phi_m \partial_z \phi_n - \phi_n \partial_z \phi_m \right|^2}{|\phi|^4} - \frac{\left|\phi_m \partial_{\bar{z}} \phi_n - \phi_n \partial_{\bar{z}} \phi_m \right|^2}{|\phi|^4} \right] \\
	{\rm tr~}\mathbf{g} &= 2\sum_{n>m}^N \left[ \frac{\left|\phi_m \partial_z \phi_n - \phi_n \partial_z \phi_m \right|^2}{|\phi|^4} + \frac{\left|\phi_m \partial_{\bar{z}} \phi_n - \phi_n \partial_{\bar{z}} \phi_m \right|^2}{|\phi|^4} \right] \\
	\mathbf{g}_{xx} - \mathbf{g}_{yy} &=  2\sum_{n>m}^N \frac{\left( \phi_m \partial_z \phi_n - \phi_n \partial_z \phi_m \right) \left( \phi_m \partial_{\bar{z}} \phi_n - \phi_n \partial_{\bar{z}} \phi_m \right)^\star + {\rm c.c.} }{|\phi|^4} \\
	\mathbf{g}_{xy} &= i\sum_{n>m}^N \frac{\left( \phi_m \partial_z \phi_n - \phi_n \partial_z \phi_m \right) \left( \phi_m \partial_{\bar{z}} \phi_n - \phi_n \partial_{\bar{z}} \phi_m \right)^\star - {\rm c.c.} }{|\phi|^4} \\
	\det \mathbf{g} &= \frac{1}{4} \left[ ({\rm tr~}\mathbf{g})^2 - (\mathbf{g}_{xx} - \mathbf{g}_{yy})^2 - 4\mathbf{g}_{xy}^2 \right]
\end{align}
One can see by inspection that the conditions $\Omega = \tr~\mathbf{g} = 2\sqrt{\det \mathbf{g}}$ and $g_{xx} - g_{yy} = g_{xy} = 0$ are satisfied if and only if $\partial_{\bar{z}} \phi_n = 0$. This constrains the unnormalized Bloch state to meromorphic functions in $k_x + i k_y$. The corresponding droplet confinement metric is just the Euclidean metric as described in the main text. Generally, one can consider an arbitrary static lattice deformation defined by transformation $\k \longrightarrow \mathbf{U}\cdot \k$. As such a transformation preserves both $\Omega$ and $\det \mathbf{g}$, $\Omega = 2\sqrt{\det \mathbf{g}}$ remains satisfied, whereas the confinement metric changes to deform the shape of the droplet accordingly.

\section{Three-Orbital $\Ch=1$ Model -- Flat Berry Curvature Optimization}

To study the effects of Berry curvature fluctuations, we introduced in the main text a $\Ch=1$ model with flat Berry curvature. This model is adiabatically connected to the three-orbital model by Sun et al. \cite{SunPRL2011} by adding longer-ranged hopping terms to minimize Berry curvature fluctuations. We start from the original lattice Hamiltonian
\begin{align}
	\Ham = \sum_\k \left[\begin{array}{lll} \CD{{\rm d},\k} & \CD{{\rm p}_x,\k} & \CD{{\rm p}_y,\k} \end{array}\right] \left[\begin{array}{lll} -2t_{dd}(\cos k_x + \cos k_y) + \delta & 2it_{pd} \sin k_x & 2it_{pd} \sin k_y \\
-2i t_{pd} \sin k_x & 2t_{pp} \cos k_x - 2t'_{pp} \cos k_y & i\Delta \\
-2i t_{pd} \sin k_y & -i\Delta & 2t_{pp} \cos k_y - 2t'_{pp} \cos k_x \end{array}\right] \left[\begin{array}{l} \C{{\rm d},\k} \\ \C{{\rm p}_x,\k} \\ \C{{\rm p}_y,\k} \end{array}\right]
\end{align}
for a three-orbital model on the square lattice with $d_{x^2-y^2}$, $p_x$, $p_y$ orbitals per site. This model exhibits a $C_4$ rotational symmetry and mirror symmetries with symmetry operators
\begin{align}
	\hat{U}_{C_4} &= \left[\begin{array}{ccc} 1 & 0 & 0 \\ 0 & 0 & 1 \\ 0 & -1 & 0 \end{array}\right] \cdot \hat{R}_{\pi/2} ~~~~~~~~~
	\hat{U}_{M} = \left[\begin{array}{ccc} 1 & 0 & 0 \\ 0 & 1 & 0 \\ 0 & 0 & -1 \end{array}\right] \cdot \hat{I}_x \cdot \hat{K}
\end{align}
where $\hat{R}_{\pi/2}$, $\hat{I}_x$ and $\hat{K}$ are $\pi/2$ rotation, x-reflection and complex conjugation operators, respectively. To proceed, we write down the most general model Hamiltonian with up to 5$^{\rm th}$ neighbor hoppings between orbitals that is invariant under the symmetry operations above. The parameter vector $\mathbf{v}$ of hopping amplitudes serve as the variational parameter of the numerical minimization of Berry curvature fluctuations using cost function
\begin{align}
	K[\mathbf{v}] = \int d^2\k \left[ \Omega(\k,\mathbf{v}) - \frac{1}{2\pi} \right]^2
\end{align}
The minimization procedure uses the original 3-orbital model parameters as starting values for $\mathbf{v}$. Furthermore, care has to be taken to ensure that the band gap remains open. We quote here the resulting model Hamiltonian that is used in the main text:
\begin{align}
	\Ham_{\rm optimized} &= \left[\begin{array}{ccc}
		h_{dd} & h_{dx} & h_{dy} \\ 
		h_{dx}^\star		& h_{xx} & h_{xy} \\
		h_{dy}^\star		& h_{xy}^\star		& h_{yy}
	 \end{array}\right]
\end{align}
with
\begin{align}
	h_{dd} &= 0.203 \cos (k_x-2 k_y)-0.0443 \cos (2 k_x-2 k_y)+2.75 \cos (k_x- \
k_y) \notag\\
	&+ 0.203 \cos (2 k_x-k_y)+2.75 \cos (k_x+k_y)+0.203 \cos (2 \
k_x+k_y)+0.203 \cos (k_x+2 k_y)-0.0443 \cos (2 k_x+2 k_y)\notag\\
	&+4.13 \cos(k_x)-0.663 \cos (2 k_x)+4.13 \cos (k_y)-0.663 \cos (2 k_y)-0.76 \\
	h_{dx} &= -(0.0372-0.1273 i) \sin (k_x-2 k_y)+(0.0391-0.0043 i) \sin (2 k_x-2 \
k_y)-(0.18-1.46 i) \sin (k_x-k_y) \notag\\
	&+(0.0439-0.0177 i) \sin (2 k_x- \
k_y)+(0.18+1.46 i) \sin (k_x+k_y)-(0.0439+0.0177 i) \sin (2 \
k_x+k_y)\notag\\
	&+(0.0372+0.1273 i) \sin (k_x+2 k_y)-(0.0391+0.0043 i) \sin (2 \
k_x+2 k_y)+1.01 i \sin (k_x)-0.374 i \sin (2 k_x)\notag\\
	&+3.10 \sin(k_y)+0.0351 \sin (2 k_y) \\
	h_{dy} &= (0.0439+0.0177 i) \sin (k_x-2 k_y)+(0.0391+0.0043 i) \sin (2 k_x-2 \
k_y)-(0.18+1.46 i) \sin (k_x- k_y)\notag\\
	&-(0.0372+0.1273 i) \sin (2 k_x- \
k_y)-(0.18-1.46 i) \sin (k_x+k_y)-(0.0372-0.1273 i) \sin (2 \
k_x+k_y)\notag\\
	&+(0.0439-0.0177 i) \sin (k_x+2 k_y)+(0.0391-0.0043 i) \sin (2 \
k_x+2 k_y)-3.10 \sin (k_x)-0.0351 \sin (2 k_x)\notag\\
	&+1.01 i \sin \
(k_y)-0.374 i \sin (2 k_y) \\
	h_{xx} &= -0.0912 \cos (k_x-2 k_y)+0.0912 \cos (2 k_x-1 k_y)+0.0912 \cos (2 \
k_x+k_y)-0.0912 \cos (k_x+2 k_y)\notag\\
	&+0.156 \cos (k_x)+0.0472 \cos (2 \
k_x)-0.156 \cos (k_y)-0.0472 \cos (2 k_y) \\
	h_{xy} &= (0.0808+0.0062 i) \cos (k_x-2 k_y)+(0.0273+0.0191 i) \cos (2 k_x-2 \
k_y)+(0.627-0.847 i) \cos (k_x-1 k_y)\notag\\
	&+(0.0808+0.0062 i) \cos (2 k_x-1 \
k_y)-(0.627+0.847 i) \cos (k_x+k_y)\notag\\
	&-(0.0808-0.0062 i) \cos (2 \
k_x+k_y)-(0.0808-0.0062 i) \cos (k_x+2 k_y)-(0.0273-0.0191 i) \cos (2 \
k_x+2 k_y)\notag\\
	&-0.921 i \cos (k_x)-0.0454 i \cos (2 k_x)-0.921 i \cos \
(k_y)-0.0454 i \cos (2 k_y)-1.07 i \\
	h_{yy} &= 0.0912 \cos (k_x-2 k_y)-0.0912 \cos (2 k_x-1 k_y)-0.0912 \cos (2 \
k_x+k_y)+0.0912 \cos (k_x+2 k_y)\notag\\
	&-0.156 \cos (k_x)-0.0472 \cos (2 \
k_x)+0.156 \cos (k_y)+0.0472 \cos (2 k_y)
\end{align}
which yields a measure of Berry curvature fluctuations
\begin{align}
	(2\pi)^2 \int d^2\k \left[ \Omega(\k,\mathbf{v}) - \frac{1}{2\pi} \right]^2 = 5.37 \times 10^{-7}
\end{align}


\begin{thebibliography}{99}
% Haldane model and Topological Insulators
\bibitem{Haldane1987}{F. D. M. Haldane, Phys. Rev. Lett. 61, 2015 (1988).}
\bibitem{HasanKaneRMP}{M. Z. Hasan, and C. L. Kane, Rev. Mod. Phys. 82, 3045 (2010).}
\bibitem{QiZhangRMP}{X. L. Qi, and S.-C. Zhang, Rev. Mod. Phys. 83, 1057 (2011).}
% Original FCI papers
\bibitem{Mudry2011}{T. Neupert, L. Santos, C. Chamon, and C. Mudry, Phys. Rev. Lett. 106, 236804 (2011).}
\bibitem{ShengNature2011}{D.N. Sheng, Z.-C. Gu, K. Sun, and L. Sheng, Nature Commun. 2, 389 (2011).}
\bibitem{RegnaultPRX2011}{N. Regnault and B. A. Bernevig, Phys. Rev. X 1, 021014 (2011).}
% Review
\bibitem{BergholtzReview2013}{E. J. Bergholtz, and Z. Liu, Int. J. Mod. Phys. B 27, 1330017 (2013).}
\bibitem{ParameswaranReview2013}{S. A. Parameswaran, R. Roy, and S. L. Sondhi, arXiv:1302.6606 (2013).}
% Physical realizations
\bibitem{XiaoOkamotoNature2011}{D. Xiao, W. Zhu, Y. Ran, N. Nagaosa, and S. Okamoto, Nat. Commun. 2, 596 (2011).}
\bibitem{VenderbosPRL2011}{J. W. F. Venderbos, S. Kourtis, J. van den Brink, and M. Daghofer, Phys. Rev. Lett. 108, 126405 (2012).}
\bibitem{KourtisPRB2012}{S. Kourtis, J. W. F. Venderbos, and M. Daghofer, Phys. Rev. B 86, 235118 (2012).}
\bibitem{CooperPRL2012}{N. R. Cooper, and R. Moessner, Phys. Rev. Lett. 109, 215302 (2012).}
\bibitem{LiuOrganometallicPRL2013}{Z. Liu, Z.-F. Wang, J.-W. Mei, Y.-S. Wu, and F. Liu, Phys. Rev. Lett. 110, 106804 (2013).}
\bibitem{GrushinPRL2014}{A. G. Grushin, \'A. G\'omez-Le\'on, and T. Neupert, Phys. Rev. Lett. 112, 156801 (2014).}
% Lattice dislocations
\bibitem{BarkeshliQiPRX2012}{M. Barkeshli, and X.-L. Qi, Phys. Rev. X 2, 031013 (2012).}
% LLL Bloch basis
\bibitem{WuRegnaultBernevigPRL2013}{Y.-L. Wu, N. Regnault, and B. A. Bernevig, Phys. Rev. Lett. 110, 106802 (2013).}
% Other ED
\bibitem{WangPRL2011}{Y.-F. Wang, Z.-C. Gu, C.-D. Gong, and D. N. Sheng, Phys. Rev. Lett. 107, 146803 (2011).}
\bibitem{WuPRB2011}{Y.-L. Wu, B. A. Bernevig, and N. Regnault, Phys. Rev. B 85, 075116 (2012).}
\bibitem{BernevigTranslationalSymmetriesPRB2012}{B. A. Bernevig, and N. Regnault, Phys. Rev. B 85, 075128 (2012).}
\bibitem{LiuRepellinPRB2013}{T. Liu, C. Repellin, B. A. Bernevig, and N. Regnault, Phys. Rev. B 87, 205136 (2013).}
\bibitem{LaeuchliPRL2013}{A. M. L\"auchli, Z. Liu, E. J. Bergholtz, and R. Moessner, Phys. Rev. Lett. 111, 126802 (2013).}
\bibitem{GrushinPRB2012}{A. G. Grushin, T. Neupert, C. Chamon, and C. Mudry, Phys. Rev. B 86, 205125 (2012).}
\bibitem{KourtisPRL2014}{S. Kourtis, T. Neupert, C. Chamon, and C. Mudry, Phys. Rev. Lett. 112, 126806 (2014).}
\bibitem{GrushinARXIV2014}{A. G. Grushin, J. Motruk, M. P. Zaletel, and F. Pollmann, arXiv:1407.6985 (2014).}
% Edge excitations
\bibitem{LiuKovrizhinPRB2012}{Z. Liu, D. L. Kovrizhin, and E. J. Bergholtz, Phys. Rev. B 88, 081106 (2013).}
\bibitem{LuoChenPRB2013}{W.-W. Luo, W.-C. Chen, Y.-F. Wang, and C.-D. Gong, Phys. Rev. B 88, 161109 (2013).}
% Haldane statistics
\bibitem{WuRegnaultBernevigPRB2014}{Y.-L. Wu, N. Regnault, and B. A. Bernevig, Phys. Rev. B 89, 155113 (2014).}
% Adiabatic continuation to FQHE
\bibitem{WuJainSunPRB2012}{Y.-H. Wu, J. K. Jain, and K. Sun, Phys. Rev. B 86, 165129 (2012).}
\bibitem{ScaffidiMoellerPRL2012}{T. Scaffidi, and G. M\"oller, Phys. Rev. Lett. 109, 246805 (2012).}
\bibitem{LiuBergholtzPRB2013}{Z. Liu, and E. J. Bergholtz, Phys. Rev. B 87, 035306 (2013).}
% Higher Chern Number ED
\bibitem{LiuPRL2012}{Z. Liu, E. J. Bergholtz, H. Fan, and A. M. L\"auchli, Phys. Rev. Lett. 109, 186805 (2012).}
\bibitem{WangYaoPRL2012}{Y.-F. Wang, H. Yao, Z.-C. Gu, C.-D. Gong, and D. N. Sheng, Phys. Rev. Lett. 108, 126805 (2012).}
\bibitem{SterdyniakPRB2013}{A. Sterdyniak, C. Repellin, B. A. Bernevig, and N. Regnault, Phys. Rev. B 87, 205137 (2013).}
\bibitem{LiuPRB2013}{Z. Liu, E. J. Bergholtz, and E. Kapit, Phys. Rev. B 88, 205101 (2013).}
\bibitem{WangYaoPRB2012}{Y.-F. Wang, H. Yao, C.-D. Gong, and D. N. Sheng, Phys. Rev. B 86, 201101 (2012).}
% Wave-Function Mappings
\bibitem{QiPRL2011}{X.-L. Qi, Phys Rev. Lett. 107, 126803 (2011).}
\bibitem{ChinghuaPRB2013}{C. H. Lee, R. Thomale, and X.-L. Qi, Phys. Rev. B 88, 035101 (2013).}
\bibitem{ChinghuaPRB2014}{C. H. Lee, and X.-L. Qi, Phys. Rev. B 90, 085103 (2014).}
\bibitem{ChaomingPRB2013}{C.-M. Jian, and X.-L. Qi, Phys. Rev. B 88, 165134 (2013).}
\bibitem{WuRegnaultBernevigPRB2012}{Y.-L. Wu, N. Regnault, and B. A. Bernevig, Phys. Rev. B 86, 085129 (2012).}
% Density operator algebra
\bibitem{ParameswaranPRB2012}{S. A. Parameswaran, R. Roy, and S. L. Sondhi, Phys. Rev. B 85, 241308(R) (2012).}
\bibitem{EstiennePRB2012}{B. Estienne, N. Regnault, and B. A. Bernevig, Phys. Rev. B 86, 241104(R) (2012).}
\bibitem{RoyARXIV2012}{R. Roy, Phys. Rev. B 90, 165139 (2014).}
\bibitem{GoerbigEPJ2012}{M. O. Goerbig, Eur. Phys. J. B 85, 15 (2012).}
\bibitem{DobardzicPRB2013}{E. Dobardzi\'c, M. V. Milovanovi\'c, and N. Regnault, Phys. Rev. B 88, 115117 	(2013).}
\bibitem{RepellinPRB2014}{C. Repellin, T. Neupert, Z. Papic, and N. Regnault, Phys. Rev. B 90, 045114 (2014).}
% Composite Fermions
\bibitem{MurthyShankarPRB2012}{G. Murthy and R. Shankar, Phys. Rev. B 86, 195146 (2012).}
%
\bibitem{RoyARXIV2014}{T. S. Jackson, G. M\"oller, and R. Roy, arXiv:1408.0843 (2014).}

% Second-quantized FQHE
\bibitem{OrtizSeidel2013}{G. Ortiz, Z. Nussinov, J. Dukelsky, and A. Seidel, Phys. Rev. B 88, 165303 (2013).}
% Lattice hamiltonians
\bibitem{SunPRL2011}{K. Sun, Z. Gu, H. Katsura, and S. Das Sarma, Phys. Rev. Lett. 106, 236803 (2011).}
\bibitem{TangWenPRL2011}{E. Tang, J.-W. Mei, and X.-G. Wen, Phys. Rev. Lett. 106, 236802 (2011).}
\bibitem{HuPRB2011}{X. Hu, M. Kargarian, and G. A. Fiete, Phys. Rev. B 84, 155116 (2011).}
% Higher Chern number
\bibitem{WangRanPRB2011}{F. Wang, and Y. Ran, Phys. Rev. B 84, 241103(R) (2011).}
\bibitem{YangHigherCPRB86}{S. Yang, Z.-C. Gu, K. Sun, and S. Das Sarma, Phys. Rev. B 86, 241112 (2012).}
\bibitem{TrescherPRB2012}{M. Trescher, and E. J. Bergholtz, Phys. Rev. B 86, 241111(R) (2012).}

% Non-Abelian phases
\bibitem{MooreRead1991}{G. Moore, and N. Read, Nucl. Phys. B 360, 362 (1991).}
\bibitem{ReadRezayi1999}{N. Read, and E. Rezayi, Phys. Rev. B 59, 8084 (1999).}
\bibitem{Ardonne1999}{E. Ardonne, and K. Schoutens, Phys. Rev. Lett. 82, 5096 (1999).}

% Anisotropic FQHE
\bibitem{HaldanePRL2011}{F. D. M. Haldane, Phys. Rev. Lett. 107, 116801 (2011).}
\bibitem{YangPRB2012}{B. Yang, Z. Papi\'c, E. H. Rezayi, R. N. Bhatt, and F. D. M. Haldane, Phys. Rev. B 85, 165318 (2012).}
\bibitem{QiuPRB2012}{R.-Z. Qiu, F. D. M. Haldane, X. Wan,  K. Yang, and S. Yi, Phys. Rev. B 85, 115308 (2012).}


% FQHE papers
\bibitem{Laughlin1983}{R. B. Laughlin, Phys. Rev. Lett. 50, 1395 (1983).}
%\bibitem{MooreRead1991}{G. Moore, and N. Read, Nucl. Phys. B 360, 362 (1991).}
%\bibitem{Halperin1983}{B. I. Halperin, Helv. Phys. Acta 56, 75 (1983).}

% Supplementary Refs
\bibitem{SUPPLEMENTARYdetailedderivation}{See Supplemental Material at [URL will be inserted by publisher], which includes Refs. \cite{SunPRL2011, DarwinFock}.}

\bibitem{DarwinFock}{V. Fock, Z. Phys. 47, 446 (1928); C. G. Darwin, Math. Proc. Cambridge Philos. Soc. 27, 86 (1930); P. A. Maksym and T. Chakraborty, Phys. Rev. Lett. 65, 108 (1990).}

% Conventional Landau levels on torus
\bibitem{HaldaneRezayi1985}{F. D. M. Haldane, and E. H. Rezayi, Phys. Rev. B 31, 2529(R) (1985).}

\bibitem{footnoteConjugateOscillatorLandauLevel}{Both the conventional and dual QH problems can be described in terms of a pair of harmonic oscillators. The FCI component index oscillator has no analogue in the conventional FQHE; conversely, FQH 'left-handed' (Landau level) degrees of freedom have no analogues in FCIs.}

\bibitem{footnoteTruncation}{The energy penalty of a steep confinement potential is approximately $\braOPket{m}{(\r/r_0)^u}{m'} \sim (\mathcal{C}/r_0)^u m^u$, hence a finite geometry with a certain filling $\nu$ may be selected by truncating at a maximum guiding center index $m_{max}$.}

\bibitem{ScaffidiPRB2014}{T. Scaffidi, and S. H. Simon, Phys. Rev. B 90, 115132 (2014).}

%\bibitem{NagaosaBook}{N. Nagaosa, Quantum Field Theory in Condensed Matter Physics}

\bibitem{footnoteGeneralTranslationOperator}{The general case follows from deformation of $e^{q_+ \omega^\mu \partial_{k_\mu}} e^{q_- \bar{\omega}^\mu \partial_{k_\mu}}$ and functions $q_{\pm}(\q)$ to satisfy $\hat{\mathcal{T}}_\q \left. \right|_{\q\to 0} \approx 1 - i q_\mu \tilde{\eta}^{\mu\nu} (-i\partial_{k_\nu})$ whilst ensuring translation symmetry of (\ref{eq:HamInteractionBare}).}

\bibitem{footnoteTwoBodyProblem}{Equivalently, two-body eigenstates of $\PID_{\rm rel} \PI_{\rm rel} \ket{\tilde{m}} = \E_{\tilde{m}} \ket{\tilde{m}}$ display perfect center-of-mass degeneracy $\PID_{\rm rel} \PI_{\rm rel} \ket{m,M} \sim m \ket{m,M}$ only for $\comm{\PI_{\rm rel}}{\PID_{\rm cm}} \to 0$.}

\bibitem{footnoteSUCsymmetry}{Weak breaking of SU($\Ch$) results from the non-commutation of left- and right-handed oscillators in the MLL problem for fluctuating magnetic field.}

\bibitem{HaldanePseudopotentials}{F. D. M. Haldane, Phys. Rev. Lett. 51, 605 (1983).}


\bibitem{ChaomingPSS}{C.-M. Jian, Z.-C. Gu, and X.-L. Qi, Phys. Stat. Sol. RRL 7, 154 (2013).}

\bibitem{Halperin1983}{B. I. Halperin, Helv. Phys. Acta 56, 75 (1983).}

\bibitem{footnoteFermionpseudopotentials}{Even-$n$ intra-component pseudopotentials $V_n$ do not contribute in the case of fermions due to Pauli exclusion.}

\bibitem{FORTHCOMING}{Details will be discussed in a forthcoming publication.}


\end{thebibliography}
\end{document}